# Accurate Performance Characterization, Reporting, and Benchmarking for Indoor Photovoltaics

Javith Mohammed Jailani[1], Amanda Luu[1], Elizabeth Salvosa[1], Charlotte Clegg[1], Vishnupriya P. Kamalon[1], Bahareh Nasrollahi[1], Irina Valitova[2], Sebastian B. Meier[3], Andrew M. Shore[4], Behrang H. Hamadani[4], Vincenzo Pecunia[1*]

[1] School of Sustainable Energy Engineering, Simon Fraser University, Surrey, V3T 0N1, BC, Canada

[2] Rayleigh Solar Tech, 1 Research Drive, Dartmouth, Nova Scotia, B2Y 4M9 Canada

[3] ASCA GmbH & Co. KG, Steigweg 24, 97318 Kitzingen, Germany

[4] National Institute of Standards and Technology, Gaithersburg, MD, 20899, USA

* Correspondence to: vincenzo_pecunia@sfu.ca (V.P.)

**ABSTRACT**

Indoor photovoltaics (IPVs) provide a promising solution for powering Internet-of-Things smart devices, which has led to a surge in IPV research. However, the diverse lighting scenarios adopted in IPV studies pose unique challenges in characterization, reporting, and benchmarking, which may obscure genuine performance improvements and result in inaccurate conclusions due to characterization errors. This study provides a comprehensive, quantitative analysis of these challenges, investigating them through the experimental characterization of IPVs covering a broad performance parameter space, including c-Si, a-Si:H, perovskite, and organic devices. We reveal that many of these challenges can lead to unacceptable error levels in IPV performance parameters, with the angular interplay among the test light source, measuring device, and IPV being particularly detrimental under diffuse indoor illumination. To address these characterization challenges, we evaluate practical protocols to mitigate them. We additionally analyze different benchmarking protocols, revealing the strengths of the reference-cell method and the limitations and solutions related to the indoor spectral coincidence concept. To facilitate the implementation of these findings, we provide comprehensive characterization, reporting, and benchmarking checklists. By enabling reliable performance evaluation and benchmarking, we anticipate that our analyses and guidelines will stimulate further advancements in IPVs, facilitating the realization of their full potential.

**MAIN**

**1. Introduction**

Indoor photovoltaics (IPVs) offer a promising solution for sustainably powering the growing number of Internet-of-Things smart devices[1,2], which has led to a surge in IPV research in recent years[3–9]. To identify and advance the most promising IPVs, accurate performance characterization, reporting, and benchmarking are crucial. However, the lack of standardized IPV performance characterization protocols prior to the recent IPV research surge has led to the adoption of a multitude of IPV characterization approaches, most of which suffer from inaccuracies and prevent benchmarking (*vide infra*). While early investigations highlighted some of these issues[10,11], comprehensive solutions have been elusive. Recently, the International Electrotechnical Commission (IEC) issued a relevant standard[12], which, however, largely focuses on general guidance rather than detailed protocols. Consequently, all experimental IPV papers published after the issuance of this standard (up to present) make no mention of it, employing characterization approaches with unconfirmed accuracy due to incomplete reporting or adopting inherently inaccurate strategies (*vide infra*). In addition to the IEC standard, dedicated test light sources (TLSs) involving sophisticated optical assemblies or system integration[13,14] have been proposed to address some of these challenges. However, these options would require IPV researchers to significantly depart from existing characterization infrastructure, necessitating considerable investments or competence in assembling these setups, thereby limiting their adoption.

Analysis of the recent literature (2023–present; see Supplementary Note 1 for details) provides insights into the continuing challenges in IPV characterization, reporting, and benchmarking. One significant challenge to accurate benchmarking is the diversity of TLSs used (Fig. 1a), as IPV performance is inherently tied to the incident spectral irradiance. In fact, fewer than 50% of these publications identify the TLS used. Characterization accuracy challenges are also prevalent. Most studies do not report aspects crucial for establishing reliable illumination conditions (Fig. 1b). Furthermore, the measuring devices (MDs) used to quantify the illuminance/irradiance are identified in only half of the works (Fig. 1c). Among the relatively few articles reporting the use of nominally calibrated luxmeters or spectroradiometers, none indicates their calibration conditions or angular responsivities (Fig. 1c). Nearly 50% of these publications rely on unspecified illumination conditions (Fig. 1d) and only 25% of them describe steps to avoid stray light affecting the device under test (DUT). Therefore, despite years of IPV research and the introduction of a relevant IEC standard, this survey underscores the urgent need to comprehensively assess the impact of characterization inaccuracies on IPV performance parameters, to gain deeper mechanistic insight into error sources, to conduct a thorough quantitative evaluation of current benchmarking protocols, and to develop any additional metrics and strategies necessary for comprehensive error assessment and mitigation.

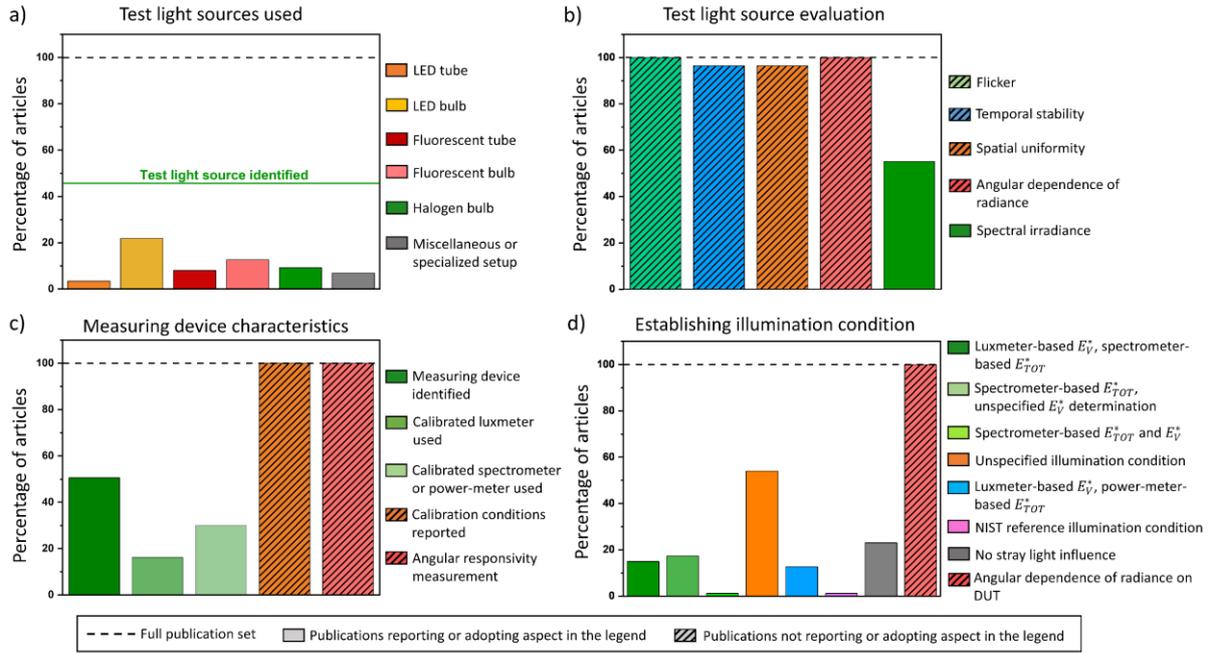

**Fig. 1| Challenges to Accurate IPV Characterization, Reporting, and Benchmarking in the Recent Literature.** These plots refer to all 87 experimental papers published from 2023 to date found on Web of Science through the key phrase 'Indoor Photovoltaics'. A solid-colored bar represents the proportion of papers reporting/adopting what is indicated in the corresponding legend. A pattern-filled bar represents the proportion of papers not reporting/adopting what is indicated in the corresponding legend. **a,** Indoor TLSs. **b,** Reported TLS characterization. **c,** MD characteristics. **d,** Illumination conditions. $E_{TOT}^*$ and $E_V^*$ are the total apparent irradiance and apparent illuminance on the DUT.

Here, we address these challenges by comprehensively examining inaccuracies specific to IPV characterization in the context of diverse IPV devices covering a broad performance parameter space. We first quantify the impact of characterization challenges on IPV performance and elucidate their origin. Building on these insights, we explore easy-to-implement mitigation strategies to ensure accurate IPV characterization using common TLSs. Furthermore, we critically assess protocols and reporting requirements to accurately benchmark IPVs, quantitatively evaluating their performance, offering new mechanistic insights into their strengths and limitations, and proposing strategies to enhance their reliability for robust interlaboratory IPV comparisons.

Before delving into our results and discussion, we note that accurate IPV characterization and reporting must also comply with requirements that are non-specific to IPV and apply to generic photovoltaic measurements (e.g., requirements related to device degradation, hysteresis and other transient effects, edge effects, shadowing, reproducibility, and temperature stability), which have been extensively covered in the photovoltaic literature[15–28] and do not warrant further analysis. Therefore, our experimental work focuses on aspects unique to the IPV context or that require a reassessment due to the specifics of IPV characterization. Nonetheless, for the convenience of IPV researchers, we also provide comprehensive checklists covering both general and IPV-specific aspects for accurate IPV characterization, reporting, and benchmarking in the Supplementary Information.

## 2. Challenges to Accurate IPV Characterization

IPV characterization can be compromised by the properties inherent to the TLSs, the methods and instrumentation used for setting the illumination condition and measuring irradiance, and the TLS-MD-DUT interplay under diffuse IPV illumination. Quantitatively assessing the IPV performance errors arising from each of these factors is essential for determining their criticality. Notably, a 10% relative error in power conversion efficiency (PCE), $\varepsilon_r^{(PCE)}$, corresponds to an absolute PCE variation, $\Delta PCE$, of ~4% for technologies achieving ~40% PCE. Since absolute PCE improvements of a few percentage points are common in research attributing such gains to materials and device engineering, thorough error analysis and reducing $\varepsilon_r^{(PCE)}$ to within 5% are essential for genuine progress.

This section provides the first comprehensive, quantitative assessment of the impact of such errors on IPV performance characterization for devices covering a wide range of performance parameters. We selected devices from three categories: those with low PCEs (labeled L), representing nascent IPV technologies; those with intermediate PCEs (labeled M1 and M2); and those with high PCEs (labeled H), representing advanced technologies. This diverse selection enables us to generalize the impact of characterization challenges and the strategies to address them.

Our experiments relied on two representative LED sources: a 30 x 30 cm² cool-white panel capped with a ground-glass diffuser (referred to as 'LED panel') and a pseudo-collimated warm-white tubular light with a 10-cm spot diameter at a 40-cm distance (referred to as 'collimated LED'). Both LEDs were equipped with regulated power supplies to stabilize the lamp output, preventing flickering—a prerequisite for accurate IPV characterization (Supplementary Fig. 1). A calibrated integrating sphere and spectroradiometer assembly (ISS)—the gold standard for spectral irradiance measurements—monitored the apparent instantaneous irradiance of the two LEDs (Supplementary Fig. 2).

## 2.1. Inaccuracies Associated with the Inherent Properties of Test Light Sources and Instrumentation

While TLS temporal stability and spatial uniformity and issues with commercial luxmeters and spectroradiometer have been identified as problematic in the IPV literature,[10,13] their quantitative impact on indoor PCE determination has remained underexplored, warranting further investigation. Our experiments reveal that TLS temporal instability and spatial non-uniformity can cause relative PCE errors well above the 5% threshold required for accurate characterization, demanding robust mitigation and thorough reporting (Supplementary Fig. 3-6, Supplementary Note 2, and Supplementary Tables 1-2). We additionally observed relative PCE errors in the region of 10% when using commercial luxmeters with unspecified reference spectra, making them unsuitable for accurate IPV characterization (Supplementary Fig. 7-8, Supplementary Table 3, and Supplementary Note 3). Spectroradiometers are therefore the strongly recommended option to characterize the irradiance and illuminance reaching the DUT, provided that they are adequately calibrated (Supplementary Note 3). Moreover, we found that the lack of compensation for dark background drift in spectroradiometer measurements can cause relative PCE errors greater than 100%, warranting robust mitigation and thorough reporting (Supplementary Fig. 7, 9, and 10; Supplementary Table 4; Supplementary Note 3).

## 2.2. IPV Characterization Inaccuracies due to Angular TLS-MD-DUT Interplay under Diffuse Indoor Illumination

Although diffuse IPV illumination is common in real-world applications and frequently used in IPV studies (Fig. 1a), it poses unique challenges to IPV characterization that have been generally overlooked in the literature and IEC 62607-7-2. This underscores the need for a thorough analysis of these challenges to gain both quantitative and mechanistic insights.

Evidence of the significance of these challenges was obtained by characterizing various IPVs illuminated by our LED panel (a highly representative diffuse TLS), placing them at two working distances ($WD_A = 12\ cm$ and $WD_B = 4\ cm$) from the panel. Meanwhile, a calibrated

ISS was employed to obtain a *nominally* identical irradiance (300 µW cm$^{-2}$) at the two working distances (Fig. 2a). Despite the nominally identical irradiance, significant deviations in the current-voltage characteristics and apparent PCE (denoted as $PCE^*$ to indicate its apparent nature) were observed for all IPVs (Fig. 2b-c). Relative deviations $\varepsilon_{r,WD}^{(PCE^*)}$ in apparent PCE between the two working distances were up to 66%, revealing an unacceptable inaccuracy level. Here, $\varepsilon_{r,WD}^{(PCE^*)}$ is defined as follows:

$$\varepsilon_{r,WD}^{(PCE^*)} \stackrel{\text{def}}{=} \frac{PCE^*(WD_B) - PCE^*(WD_A)}{PCE^*(WD_A)} \qquad (1)$$

We trace these deviations to the TLS-MD-DUT angular interplay under diffuse IPV illumination. The responsivity of any MD/DUT generally varies with the angle of incidence: $R_{MD} = R_{MD}(\theta, \varphi)$ and $R_{DUT} = R_{DUT}(\theta, \varphi)$, respectively (where $\theta$ and $\varphi$ are the angular coordinates illustrated in Fig. 2a; see also Supplementary Note 4). Generally, $R_{MD}(\theta, \varphi)$ and $R_{DUT}(\theta, \varphi)$ differ from the corresponding responsivities under normal illumination ($R_{MD,\perp}$ and $R_{DUT,\perp}$, respectively). Moreover, as the distance between an MD (or DUT) and a diffuse TLS changes, the angular distribution of the incoming radiance $L$ also varies. Therefore, for a DUT-MD pair with different normalized angular responsivities (i.e., $R_{MD}(\theta, \varphi)/R_{MD,\perp} \neq R_{DUT}(\theta, \varphi)/R_{DUT,\perp}$), the MD may detect the same apparent irradiance $E^*_{TOT}$ at two working distances upon suitably adjusting the TLS (as in Fig. 2) while the DUT senses different amounts of light. This causes the DUT to manifest different photovoltaic parameters at different working distances under the same nominal irradiance (as per MD) compared to a truly unchanged illumination condition, leading to inaccuracies in the determination of its PCE.

The relative error in apparent PCE under the experimental conditions presented in Fig. 2a can be expressed as:

$$\begin{aligned}\Delta PCE^*_{B-A} &= PCE^*(WD_B) - PCE^*(WD_A) \\ &\cong \frac{V_{OC}\, FF}{E^*_{TOT}} \iint_{\Omega^-} R_{DUT,\perp} \left( \frac{R_{DUT}(\theta, \varphi)}{R_{DUT,\perp}} - \frac{R_{MD}(\theta, \varphi)}{R_{MD,\perp}} \right) \Delta L^{(B-A)}(\theta, \varphi) \sin\theta\, d\theta\, d\varphi \end{aligned} \quad (2)$$

(Supplementary Notes 4-9). Here, $\Delta L^{(B-A)} = L^{(B)}(\theta, \varphi) - L^{(A)}(\theta, \varphi)$ defines the difference in radiance at the two working distances (i.e., $L^{(B)}(\theta, \varphi)$ and $L^{(A)}(\theta, \varphi)$), and $V_{OC}$ and $FF$ are the DUT's open-circuit voltage and fill factor. The integration is carried out over the region $\Omega^-$ consisting of the hemisphere above the DUT minus the region in the vicinity of the DUT's surface normal.

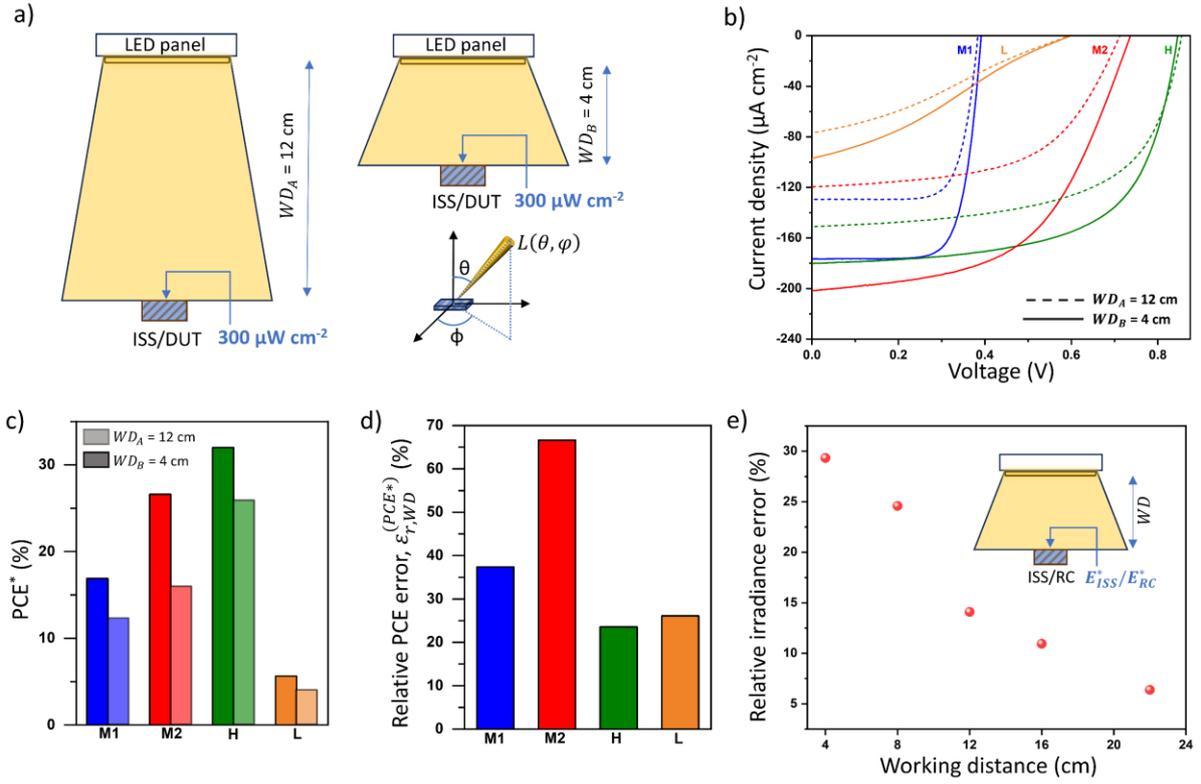

**Fig. 2| Inaccuracies due to the TLS-MD-DUT Angular Interplay under Diffuse IPV Illumination. a,** DUT/ISS placed at two working distances ($WD_A = 12\ cm$ and $WD_B = 4\ cm$) from an LED panel, which was adjusted to an irradiance of 300 µW cm$^{-2}$ (as per ISS; Supplementary Note 10) at both distances. The spherical coordinate system for the radiance $L = L(\theta, \varphi)$ is also shown. **b,** Current density-voltage (J-V) characteristics and **c,** corresponding apparent PCEs for various DUTs under the same apparent irradiance (300 µW cm$^{-2}$) at the two distances. **d,** Corresponding relative PCE variations $\varepsilon_{r,WD}^{(PCE^*)}$. **e,** Relative irradiance error as a function of distance between the calibrated ISS, whose angular responsivity significantly departs from $\cos\theta$, and another calibrated device (labeled RC) with normalized angular responsivity closer to $\cos\theta$. $E_{ISS}^*$ and $E_{RC}^*$ indicate the apparent irradiance values measured by the two MDs.

The TLS-MD angular interplay under diffuse IPV illumination can also increase the PCE error due to inaccurate irradiance determination. If $R_{MD}(\theta, \varphi)$ deviates from $\cos\theta$, the MD will inaccurately weigh radiance from different angles, introducing an irradiance error given by:

$$\Delta E_{TOT} = \iint_{\Omega^-} \left( \cos\theta - \frac{R_{MD}(\theta,\varphi)}{R_{MD,\perp}} \right) L(\theta,\varphi) \sin\theta\ d\theta\ d\varphi \qquad (3)$$

(Supplementary Note 7). Eq. 2-3 and discussion thereof justify the label 'apparent PCE' for the PCE determined by dividing the DUT's maximum power density and the irradiance determined by a calibrated MD, as it is generally not the true PCE.

Based on Eq. 2-3, accurate IPV characterization under a diffuse indoor light source firstly requires that the DUT's and MD's normalized responsivities have the same angular dependence, ensuring that both devices respond to the same amount of light. Additionally, the MD should approach a $\cos\theta$-compliant behaviour to accurately measure irradiance. In practice, these conditions do not need to hold for all angles, but only within the angular range over which the radiance on the IPV is significant (Eq. 2-3).

Further evidence supporting our interpretation of Fig. 2a-d comes from angular responsivity measurements (Supplementary Fig. 11). We confirmed that our calibrated ISS's angular responsivity is appreciably narrower than $\cos\theta$, justifying its inaccurate irradiance determination under diffuse IPV illumination (Eq. 3). We further verified this effect by comparing the irradiance measured by the calibrated ISS against a calibrated MD (labeled RC) with an angular responsivity closer to $\cos\theta$ behavior, finding that the ISS suffered from an irradiance inaccuracy > 30% at low working distances (Fig. 2e), where $L(\theta,\varphi)$ is broader. Furthermore, we found that the ISS's angular responsivity is narrower than that of all DUTs considered, explaining the pronounced variations in their apparent PCEs with working distance (Fig. 2). Our angular responsivity data also aligns with the trend in Fig. 2d, where devices with the largest errors in short-circuit current density ($J_{SC}$) and PCE deviate more from the ISS's angular response.

We note that the observed working-distance issue is not specific to our ISS. We found similar issues when using a commercial optical power meter (OPM) with NIST-traceable calibration as the MD (Supplementary Fig. 12). This highlights the problems inherent in characterizing IPVs using instrumentation designed and calibrated for normal collimated illumination under diffuse IPV illumination. Furthermore, even if a $\cos\theta$-compliant MD is used, Eq. 2-3 indicate that the working-distance effect is generally not overcome due to the interplay between the angular behavior of the TLS and IPV device (Supplementary Notes 8-9).

### 3. Protocols for Improved Accuracy under Diffuse IPV Illumination
Herein, we discuss strategies to mitigate the IPV characterization inaccuracies due to the MD-DUT-TLS angular interplay under diffuse IPV illumination.

## 3.1. Dark-Tube Method

If the conditions for accurate IPV characterization (i.e., $\cos\theta$-compliant MD and $R_{MD}(\theta, \varphi)/R_{MD,\perp} \cong R_{DUT}(\theta, \varphi)/R_{DUT,\perp}$ within the TLS's angular range) are not met for a given diffuse indoor TLS, an easy-to-adopt approach to considerably improve accuracy is to restrict the $\theta$ range over which $L(\theta, \varphi)$ is significant. This ensures that the integrals in Eq. 2-3 are restricted to a $\theta$ range such that $R_{MD}(\theta, \varphi) \cong R_{MD,\perp}$ and $R_{DUT}(\theta, \varphi) \cong R_{DUT,\perp}$, making $\varepsilon_{r,WD}^{(PCE^*)}$ and $\Delta E_T$ negligible.

We validated this approach by illuminating our MDs/DUTs with our LED panel through an optically absorbing tube. Using tubes of different lengths for varying working distances (Fig. 3a), this approach considerably reduced discrepancies in IPV performance as the working distance changed (Fig. 3b-c). The apparent PCE error $\varepsilon_{r,WD}^{(PCE^*)}$ under global illumination (up to 43%) was reduced by up to nearly an order of magnitude with the dark-tube approach, narrowing the relative error range to approximately 5% to 7% (Fig. 3c). Further error reduction could potentially be achieved by narrowing the tube's cross section (Supplementary Fig. 13). Despite its effectiveness, this approach has a practical limitation: only a small fraction of the light output from the diffuse TLS is utilized, requiring much higher electrical power to drive the TLS so that the light output through the tube is comparable to the optically unconstrained TLS case (e.g., an order-of-magnitude increase in electrical power was required for our diffuse indoor TLS). Therefore, due to the operational limitations of typical TLSs for indoor lighting, this method is feasible only for low illuminances, preventing its use for characterization up to 1000 lx.

## 3.2. Reference-Cell Method for IPV Characterization

Calibrated reference cells (RCs) are well established in solar photovoltaics for characterization with minimal spectral mismatch errors. NIST has developed a calibrated RC for IPV benchmarking[11], enabling the DUT's photovoltaic characteristics to be mapped to three 1000-lx standard reporting conditions (Fig. 4a; Supplementary Note 11). While the RC method aims to address inaccuracies due to spectral mismatch, it could potentially be a powerful tool for characterizing IPVs under diffuse indoor TLSs, provided that the selected RC meets the conditions for accurate characterization discussed in Section 2.2. To assess the viability of this approach, we used an RC calibrated according to the NIST protocol to characterize our IPVs under diffuse indoor illumination (as provided by our LED panel) at 4 cm and 12 cm distances (Fig. 4b), mapping their characteristics to NIST's CCT3000K standard reporting condition (Supplementary Note 11). The resulting current-voltage characteristics for different working distances revealed minimal discrepancies, with $\varepsilon_{r,WD}^{(PCE^*)} \leq 3.1\%$ (Fig. 4c-d). For comparison, measurements using our calibrated ISS gave $\varepsilon_{r,WD}^{(PCE^*)}$ of up to 35% for the same illumination conditions (Fig. 4d).

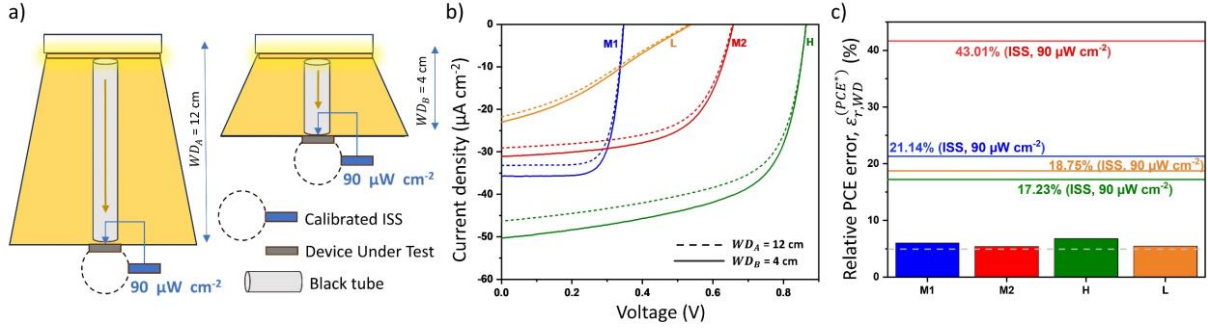

**Fig. 3| Dark-Tube Method. a,** Experimental setup comprising an LED panel and a dark tube connecting it to the DUT/ISS. The irradiance through the tube was set to 90 μW cm$^{-2}$ (as per ISS; Supplementary Note 10) at 4 cm and 12 cm distances. **b,** Corresponding J-V characteristics of various DUTs. **c,** Relative PCE deviations $\varepsilon_{r,WD}^{(PCE^*)}$ between the two distances. Solid lines represent the $\varepsilon_{r,WD}^{(PCE^*)}$ values obtained under equivalent global illumination (i.e., without the tube) (Supplementary Fig. 14).

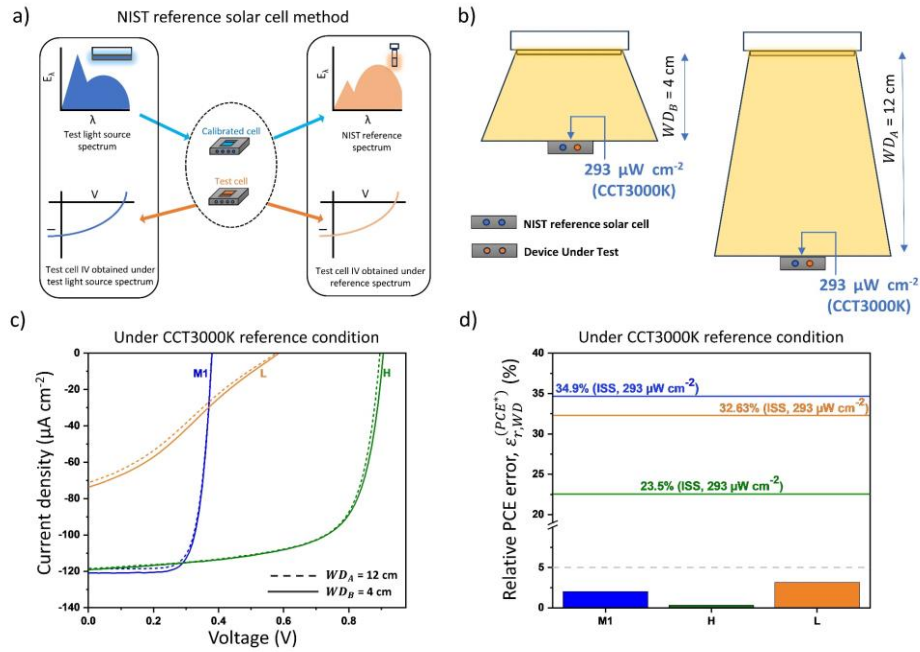

**Fig. 4| Reference-Cell Method under Diffuse IPV Illumination. a,** Concept underlying the RC method for IPV characterization. **b,** Experimental setup with an LED panel and DUT/RC. The indoor TLS was adjusted to achieve the NIST CCT3000K standard reference condition (293 μW cm$^{-2}$, 1000 lx) at 4 cm and 12 cm distances (as per RC). **c,** Corresponding J-V characteristics of various DUTs at these distances. **d,** Relative PCE deviations $\varepsilon_{r,WD}^{(PCE^*)}$ between the two distances. Solid lines represent the $\varepsilon_{r,WD}^{(PCE^*)}$ values obtained at the two distances when the LED panel was adjusted to 293 μW cm$^{-2}$, as per calibrated ISS.

We trace the minimal $\varepsilon_{r,WD}^{(PCE^*)}$ obtained with the RC method to the RC's and DUT's similar angular responses (Supplementary Fig. 11), consistent with Eq. 2. Note that, while achieving a point-by-point identity between $R_{MD}(\theta, \varphi)/R_{MD,\perp}$ and $R_{DUT}(\theta, \varphi)/R_{DUT,\perp}$ is unrealistic, the invariance of a DUT's photovoltaic characteristics with working distance (as obtained with the RC method) provides a solid indication of highly similar $R_{MD}(\theta, \varphi)/R_{MD,\perp}$ and $R_{DUT}(\theta, \varphi)/R_{DUT,\perp}$ on average (with an accuracy given by $\varepsilon_{r,WD}^{(PCE^*)}$; see Eq. 2). However, minimal $\varepsilon_{r,WD}^{(PCE^*)}$ values, in principle, do not resolve the 'apparent' nature of the PCE. Indeed, RCs are typically calibrated under normal illumination, hence the irradiance associated with PCE calculations based on the RC method may be subject to errors under diffuse IPV illumination (Eq. 3). This aspect is expanded upon in Section 5.2, where we validate the accuracy of RC-determined PCEs under our diffuse IPV illumination condition.

Unlike the dark-tube approach, the RC method allows characterization over the entire IPV illuminance range (50 lx to 1000 lx) using typical indoor lamps—with the added benchmarking benefit discussed in Section 5.2—considerably enhancing its deployability. However, the RC method requires knowledge of the DUT's spectral responsivity, $R_{DUT}(\lambda)$, and ensuring the similarity of the RC's and DUT's angular responses. While measuring $R_{DUT}(\lambda)$ for IPV cells and parallel-connected IPV modules is straightforward in most laboratories, challenges arise with series-connected modules, which require specialized characterization systems.

### 4. Accurate Characterization under Normal Collimated IPV Illumination

The angular TLS-MD-DUT interplay ceases to be an issue under normal collimated IPV illumination, which also resolves the working-distance effect (Supplementary Notes 7-10 and 12 and Supplementary Fig. 15). Therefore, accurate IPV measurements can be conducted using a collimated indoor TLS for normal illumination. Irradiance and illuminance should be determined using an ISS calibrated under normal illumination by an accredited institution (Supplementary Note 10). Accurate dark background subtraction for the ISS remains crucial (Supplementary Note 3). Importantly, this approach can be easily scaled to large-area IPVs (Supplementary Fig. 16).

## 5. Accurate IPV Benchmarking

Even if all the accuracy challenges discussed so far are duly addressed, seamless benchmarking of IPV performance will still not be possible due to the diverse range of TLSs used in IPV characterization. Herein, we critically examine different benchmarking strategies, providing quantitative insights into their general feasibility and identifying additional metrics needed for accurate implementation.

### 5.1. Challenges with CCT and Indoor Spectral Coincidence

A widely pursued approach to IPV benchmarking has focused on standardizing the TLS. However, our analysis reveals challenges that have been overlooked, necessitating refinements and the introduction of additional metrics for accuracy.

First, for non-collimated indoor TLSs, even with complete spectral overlap with a standard reference irradiance spectrum, accurate benchmarking is unattainable due to large errors associated with diffuse IPV illumination and the difficulties in mitigating them (Section 2.2). This issue, not addressed in IEC 62607-7-2, is critically important for benchmarking because most commercial indoor TLSs are non-collimated. Therefore, IPV benchmarking through indoor TLS standardization must rely on collimated or pseudo-collimated indoor light sources.

With collimated indoor TLSs, the challenge of accurate benchmarking via TLS standardization shifts to achieving spectral overlap with a standard reference irradiance spectrum. Since no indoor TLS can perfectly match a standard spectral irradiance, metrics are needed to assess how closely a TLS irradiance spectrum aligns with the chosen reference.

IPV applications operate under various illumination conditions, typically described by their correlated color temperatures (CCTs), which can range from 3000 K to 6500 K. However, in the following, we reveal that CCT alone is an incomplete descriptor for accurate IPV benchmarking, even within specific indoor TLS categories (e.g., phosphor-based white LEDs or fluorescent sources). Firstly, indoor TLSs may deviate significantly (e.g., by ± 300 K) from their stated CCTs. For instance, Fig. 5a shows that five commercial white LEDs labeled as 4000 K[29–33] actually range from 3863 K to 4362 K, as determined from their spectral irradiance (Supplementary Note 13). Such variations inevitably lead to large discrepancies in total

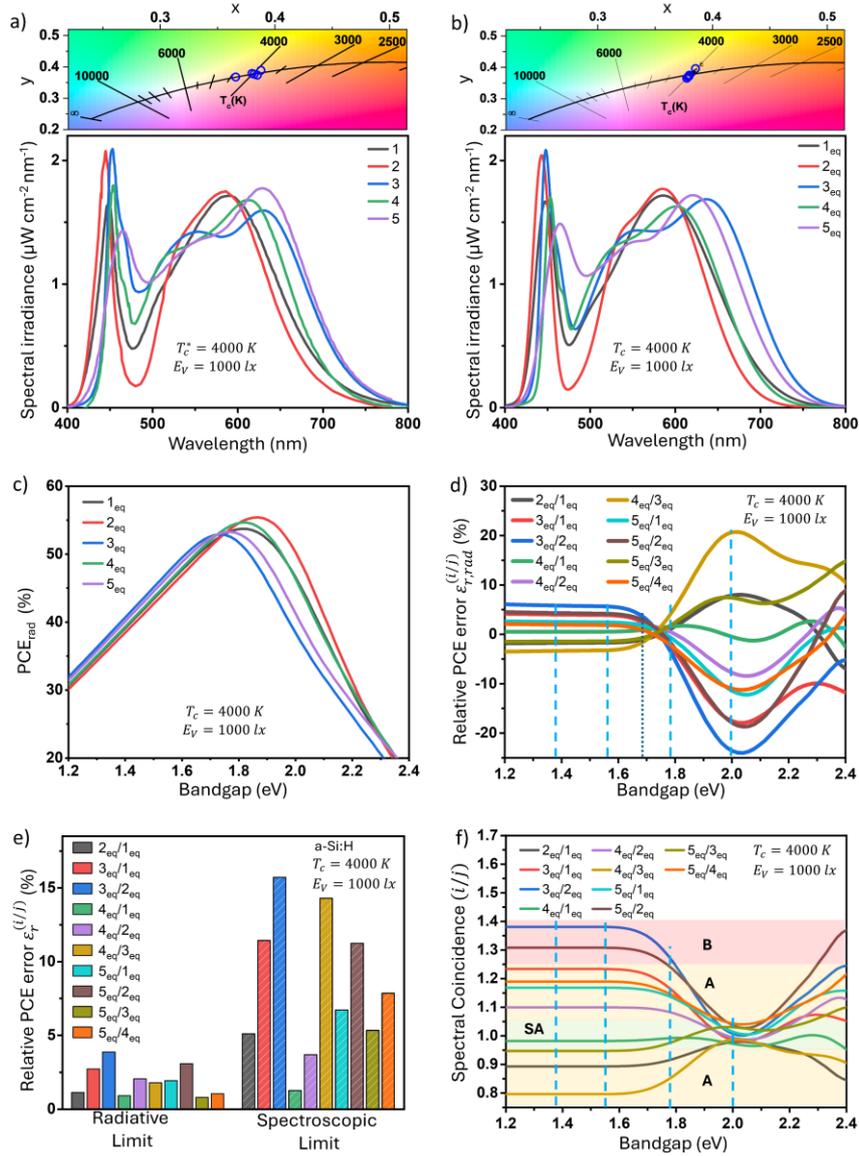

**Fig. 5| Benchmarking Challenges with CCT and Indoor Spectral Coincidence. a,** Irradiance spectra of five commercial indoor TLSs[29–33] with a nominal 4000 K CCT and 1000 lx illuminance, alongside their placement in Judd's uniform chromaticity space (blue circles). **b,** Irradiance spectra with exact 4000 K CCT and 1000 lx illuminance, derived from the spectra in Fig. 5a via a Gaussian-based reconstruction algorithm, alongside their placement in Judd's uniform chromaticity space (blue circles). **c,** Radiative-limit PCE for the spectra in Fig. 5b. **d,** Relative radiative-limit PCE discrepancies derived from Fig. 5c. The dotted line corresponds to the a-Si:H IPV referred to in Fig. 5e. **e,** Relative PCE discrepancies in the radiative and spectroscopic limits for a model a-Si:H IPV[34] illuminated with the irradiance spectra in Fig. 5b. **f,** Generalized indoor spectral coincidence corresponding to the irradiance spectra in Fig. 5b. The dashed lines in Fig. 5d and 5f mark the energy boundaries associated with the indoor spectral coincidence definition in IEC 62607-7-2. Shaded regions and associated labels denote the corresponding indoor spectral coincidence classes, as defined in IEC 62607-7-2.

− 13 −

irradiance, as well as in PCE for a given DUT (Supplementary Fig. 17-18). Therefore, relying solely on CCT for indoor TLS standardization, even within a specific indoor TLS category, creates accuracy challenges, firstly because commercial lighting manufacturers typically do not implement spectral requirements as stringent as those needed for accurate IPV benchmarking.

The benchmarking accuracy issue persists even with indoor TLSs of identical CCT within the same lamp category. To demonstrate this, we applied a Gaussian-based spectrum reconstruction algorithm with parameter optimization to generate irradiance spectra (Fig. 5b) at exactly 4000 K, closely resembling the commercial LEDs in Fig. 5a (see also Supplementary Fig. 18). We then calculated the corresponding single-junction PCEs in the radiative limit ($PCE_{rad}$) for a continuum of DUTs with an ideal external quantum efficiency (EQE) of 100% for above-bandgap photon energies (Supplementary Note 14). The resulting $PCE_{rad}^{(i)}$ traces (where $i$ denotes the $i$-th spectrum in Fig. 5b) highlight significant discrepancies for bandgap energies relevant to IPVs, i.e., from 1.8 eV to 2.0 eV (Fig. 5c). Although appreciable $PCE_{rad}$ differences for varying CCTs have been reported[35,36], similarly large discrepancies for indoor TLSs with identical CCT within the same lamp category had not been reported before. Fig. 5d quantifies these discrepancies as $\left(p_{max,rad}^{(i)}(E_g)/E_{TOT}^{(i)} - p_{max,rad}^{(j)}(E_g)/E_{TOT}^{(j)}\right)/\left(p_{max,rad}^{(j)}(E_g)/E_{TOT}^{(j)}\right)$, where $i$ and $j$ denote the indoor TLS and reference spectra, respectively, and $p_{max,rad}^{(i)}(E_g)$ is the maximum output power density in the radiative limit for a single-junction device with absorber's bandgap $E_g$ under illumination from the $i$-th indoor TLS with total irradiance $E_{TOT}^{(i)}$. This reveals unacceptable discrepancies exceeding the 5% threshold required for benchmarking accuracy for nearly all spectral pairs, with some surpassing 20%.

To assess benchmarking accuracy for a DUT with a non-constant EQE for above-bandgap photon energies, we simulated an a-Si:H IPV[34] ($E_g = 1.67$ eV) in the spectroscopic limit—i.e., incorporating the actual EQE spectrum into the photovoltaic performance calculations (Supplementary Note 14). Although a bandgap of 1.67 eV leads to minor $PCE$ discrepancies in the radiative limit (Fig. 5e and dotted line in Fig. 5d), discrepancies for most spectral pairs exceed the 5% accuracy threshold and can reach up to 15% in the more realistic spectroscopic limit. This indicates that PCE discrepancies do not only depend on the reference and test irradiance spectra; they may also be amplified by the way a DUT's spectral response interacts with the differences between reference and test irradiance spectra. This effect varies across spectral combinations and cannot be captured by a simple multiplicative factor. Overall, our analysis reveals, for the first time, that relying solely on CCT for TLS selection, even with TLSs of identical CCT within the same lamp category, may results in significant benchmarking inaccuracies.

To mitigate this challenge, IEC 62607-7-2 introduced the indoor spectral coincidence concept to assess how closely an indoor TLS matches a standard reference spectrum. The indoor spectral coincidence $ISC$ of a TLS with spectral irradiance $E_{TLS}(\lambda)$ relative to a reference spectral irradiance $E_{ref}(\lambda)$ is defined as:

$$ISC \stackrel{\text{def}}{=} \frac{\int_\Lambda R_{\bar{k}}(\lambda) E_{TLS}(\lambda) d\lambda}{\int_\Lambda R_{\bar{k}}(\lambda) E_{ref}(\lambda) d\lambda} \tag{4a}$$

where $R_k(\lambda)$ is the spectral responsivity of the $k$-th reference device and $\bar{k}$ is given by the formula

$$\bar{k} \stackrel{\text{def}}{=} \operatorname{argmax}_k \left| \frac{\int_\Lambda R_k(\lambda) E_{TLS}(\lambda) d\lambda}{\int_\Lambda R_k(\lambda) E_{ref}(\lambda) d\lambda} - 1 \right| \tag{4b}$$

Note that the five reference devices identified in IEC 62607-7-2 have nearly flat EQE spectra for photon energies above 2 eV, 1.77 eV, 1.55 eV, 1.37 eV, and 1.12 eV, respectively, within the visible range. From an $ISC$ standpoint, IEC 62607-7-2 classifies a TLS as class-SA if $|ISC - 1| \leq 0.05$, class-A if $|ISC - 1| \leq 0.25$, and so forth.

While $ISC$ reflects *short-circuit current* ratios for the reference devices (Eq. 4), IPV benchmarking typically involves the DUT's *PCE* (or related quantities such as the maximum output power density). To examine how the $ISC$ concept applies to robust benchmarking from a PCE perspective, we define a generalized indoor spectral coincidence $\widetilde{ISC}(E_g; i, j)$ as a function of bandgap:

$$\widetilde{ISC}(E_g; i, j) \stackrel{\text{def}}{=} \frac{\int_\Lambda R_{E_g}(\lambda) E_i(\lambda) d\lambda}{\int_\Lambda R_{E_g}(\lambda) E_j(\lambda) d\lambda} \tag{5}$$

Here, $R_{E_g}(\lambda)$ is the spectral responsivity of an ideal device with an EQE of 100% for photon energies above $E_g$, and $E_i(\lambda)$ and $E_j(\lambda)$ refer to the irradiance spectra designated as TLS and reference, respectively. This definition leverages the similarity between the test devices identified in IEC 62607-7-2 and the assumption of a constant EQE for above-bandgap photon energies inherent in radiative-limit PCE calculations.

We evaluated $\widetilde{ISC}(E_g; i, j)$ for all permutations of the five 4000 K irradiance spectra in Fig. 5b. The corresponding $\widetilde{ISC}(E_g; i, j)$ traces (Fig. 5f) reveal that, from an $ISC$ standpoint, only two spectral pairs are class-SA, many others fall into class A, and two are within class B, following the IEC 62607-7-2 approach.

By comparing $\widetilde{ISC}$ against the corresponding PCE discrepancies in Fig. 5d, we find that class-SA spectral pairs (e.g., the one labeled 5$_{eq}$/3$_{eq}$ in Fig. 5) can result in PCE discrepancies exceeding the 5% threshold required for robust benchmarking within the IPV target bandgap range. Class-A pairs (e.g., the ones labeled 3$_{eq}$/1$_{eq}$, 5$_{eq}$/4$_{eq}$, 5$_{eq}$/1$_{eq}$, 2$_{eq}$/1$_{eq}$, 4$_{eq}$/3$_{eq}$ in Fig. 5) lead to even larger PCE discrepancies of up to 20% within the same bandgap range. As previously discussed, the spectral responsivities of actual DUTs may exacerbate PCE discrepancies compared to the radiative limit (e.g., see Fig. 5e). Given that remaining within a 5% discrepancy threshold is critical for benchmarking, it is evident that an indoor TLS with a class-A $ISC$ does not generally guarantee benchmarking accuracy. While indoor TLSs with a class-SA $ISC$ offer better spectral robustness, they may still result in inaccurate benchmarking depending on the DUT's bandgap and spectral responsivity.

The core issue with this benchmarking approach is that the PCE benchmarking discrepancy is specific to the IPV measurement and DUT at hand, rather than being a mere TLS property. For a given reference spectrum and TLS, the PCE benchmarking discrepancy $\varepsilon_{r,bm}^{(PCE)}$ can be expressed as:

$$\varepsilon_{r,bm}^{(PCE)} = \frac{PCE_{TLS} - PCE_{ref}}{PCE_{ref}} \tag{6a}$$

$$= \frac{J_{SC,TLS}}{J_{SC,ref}} \frac{E_{ref}}{E_{TLS}} \frac{V_{OC,TLS}}{V_{OC,ref}} \frac{FF_{TLS}}{FF_{ref}} - 1 \tag{6b}$$

Here, the DUT's PCE under indoor TLS illumination, $PCE_{TLS}$, is expressed as $J_{SC,TLS} V_{OC,TLS} FF_{TLS} / E_{TOT,TLS}$ and analogous quantities are used for the PCE under reference illumination, $PCE_{ref}$. The ratio $J_{SC,TLS}/J_{SC,ref}$ in Eq. 6b is formally similar to the $ISC$ as defined in IEC 62607-7-2 (Eq. 4), given that:

$$\frac{J_{SC,TLS}}{J_{SC,ref}} = \frac{\int_\Lambda R_{DUT}(\lambda, E_{TLS}(\lambda)) \, E_{TLS}(\lambda) \, d\lambda}{\int_\Lambda R_{DUT}(\lambda, E_{ref}(\lambda)) \, E_{ref}(\lambda) \, d\lambda} \tag{7}$$

Here, $R_{DUT}$ is the spectral responsivity of the entire DUT (rather than a mere portion of it). In fact, Eq. 7 significantly differs from the $ISC$ definition, given the distinct dependence of the generic DUT's responsivity on wavelength and incident spectral irradiance compared to the reference devices utilized in calculating $ISC$. Such differences, alongside differences in open-circuit voltage and fill factor under reference and TLS illumination and effects from parasitic resistances (Eq. 6), confirm that $ISC$ does not provide a direct measure of the PCE discrepancy $\varepsilon_{r,bm}^{(PCE)}$, consistent with Fig. 5.

Accuracy within this benchmarking approach can only be confirmed by ensuring that $\varepsilon_{r,bm,est}^{(PCE)}$, which we introduce as a new benchmarking metric based on the definition below, is ≤5%:

$$\varepsilon_{r,bm,est}^{(PCE)} \stackrel{\text{def}}{=} \frac{p_{max,TLS}}{p_{max,ref,est}} \frac{E_{TOT,ref}}{E_{TOT,TLS}} - 1 \qquad (8)$$

Here, $p_{max,TLS}$ and $p_{max,ref,est}$ are the DUT's measured maximum output power density under TLS illumination and estimated maximum output power density under standard reference illumination, respectively, set to the same target illuminance, and $E_{TOT,TLS}$ and $E_{TOT,ref}$ are the measured TLS's total irradiance and the total standard reference irradiance at the target illuminance, respectively. $p_{max,ref,est}$ should be determined by calculating the maximum power point from the J-V characteristic $J_{ref,est}(V) = J_{TLS}(V) + \int_\Lambda R_{DUT}(\lambda, E_{TLS}(\lambda)) \left[ E_{TLS}(\lambda) - E_{ref}(\lambda) \right] d\lambda$, where $J_{TLS}(V)$ is the measured J-V characteristic under TLS illumination at the target illuminance (Supplementary Note 15, Supplementary Fig. 19). In contrast to $ISC$, $\varepsilon_{r,bm,est}^{(PCE)}$ quantifies benchmarking accuracy specific to the DUT and indoor TLS used (Supplementary Note 15, Supplementary Fig. 19), which is crucial for ensuring reliable interlaboratory comparisons.

We note that IEC 62607-7-2 defines PCE in a slightly different manner in the context of $ISC$-based benchmarking from direct current-voltage measurements at the target illuminance ("illuminometer method") (Supplementary Note 16). However, our analysis reveals that TLSs with class-SA or class-A $ISC$s do not guarantee accurate benchmarking also under this alternative definition, requiring the introduction of a new metric, analogous to $\varepsilon_{r,bm,est.}^{(PCE)}$, for accurate reporting and benchmarking (Supplementary Note 16; Supplementary Fig. 20).

### 5.2. IPV Benchmarking via the Reference-Cell Method

Departing from hard requirements on the TLS's spectral irradiance, NIST adapted the RC method for IPV characterization[11], introducing three reference spectral irradiances (CCT3000K, CCT4000K, and CCT6000K; Fig. 6a) as standard reporting conditions to which IPV performance can be mapped (Fig. 4a; Supplementary Note 11). With one single RC and one single indoor TLS, traceable PCEs for three different spectral conditions can be determined with an accuracy determined by the accuracy of the calibration itself and the current-voltage measurement equipment adopted, both typically within 2%. Several years since its publication, however, this method has hardly been used in IPV research (Fig. 1). Thus, the accurate benchmarking problem has remained unresolved.

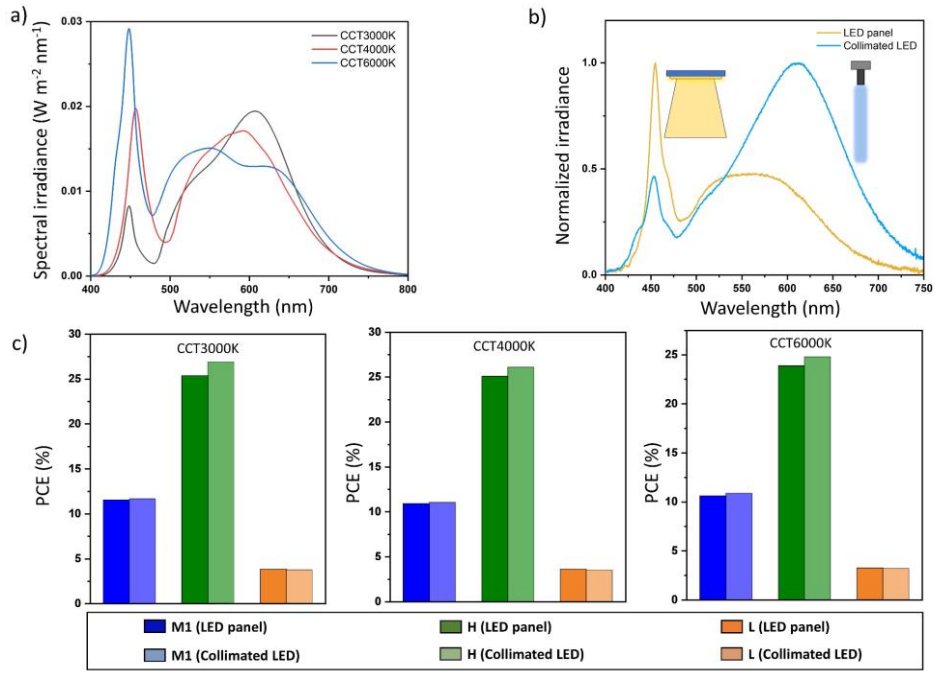

**Fig. 6| Benchmarking via NIST Reference Cell for IPV Characterization. a,** Spectral irradiance for the NIST standard reporting conditions (1000 lx). **b,** Normalized irradiance for the two indoor TLSs used during the application of the RC method to our IPVs. **c,** RC-derived PCEs for our DUTs under the three NIST standard reporting conditions for IPVs.

To explore the versatility of this method and its performance with various indoor TLS conditions and IPV technologies, we applied it to our diverse IPVs (i.e., M1, H, and L cells; Supplementary Fig. 21 and Supplementary Table 5) under all NIST standard reporting conditions, using both our diffuse LED panel (CCT6500K) and collimated LED (CCT3000K) (Fig. 6b).

The RC-method-derived performance of each of the IPVs was nearly unaffected by the TLS used (regardless of its CCT and direct/diffuse character) for all standard reporting conditions (Fig. 6c). Note that PCE differences across different standard reporting conditions do not indicate accuracy issues, as they reflect inherent IPV performance variations depending on the incident spectral irradiance. For each standard reporting condition, the relative PCE discrepancy $\varepsilon_{r,TLS}^{(PCE^*,RC)}$ (Eq. 9) was within 5% for all IPVs as the indoor TLS was varied.

$$\varepsilon_{r,TLS}^{(PCE^*,RC)} = \frac{PCE_{RC}^*(Diffuse\ LED\ Panel) - PCE_{RC}(Collimated\ LED)}{PCE_{RC}(Collimated\ LED)} \quad (9)$$

In Eq. 9, we use $PCE_{RC}(Collimated\ LED)$ without an asterisk to indicate the PCE accuracy under collimated LED illumination. This is because the RC used was calibrated under normal illumination, replicating the normal illumination condition realized with our collimated LED. Therefore, inaccuracies associated with diffuse IPV illumination (Eq. 1-3) are absent in $PCE_{RC}(Collimated\ LED)$. This consideration provides insight into the 'apparent' nature of the PCE of our devices under diffuse IPV illumination ($PCE_{RC}^*(Diffuse\ LED\ Panel)$), as determined with the NIST RC. An $\varepsilon_{r,TLS}^{(PCE^*,RC)}$ within 5% across all IPVs and illumination conditions considered implies that $PCE_{RC}^*(Diffuse\ LED\ Panel)$ is practically indistinguishable from $PCE_{RC}(Collimated\ LED)$ within the accuracy levels achievable in IPV research laboratories (provided that all other fundamental accuracy challenges discussed earlier are resolved). Therefore, while the NIST-traceable RC adopted was calibrated under normal illumination, Fig. 6c indicates that the associated error in total irradiance (Eq. 3) was minimal under illumination from our diffuse LED panel for the experimental conditions used. This demonstrates the versatility of the RC method for IPV characterization when RCs are used that minimally deviate from $\cos\theta$ behaviour within the angular range where the incident radiance is significant under the experimental conditions used.

### 5.3. Illuminance and Irradiance Levels for Benchmarking

Accurately measured irradiance and illuminance are significant for IPV benchmarking (Supplementary Note 17) only when referenced to a standard spectral irradiance, either through an ISC- or RC-based method. The specific illuminance/irradiance levels to be adopted are a matter of convention rather than accuracy, provided that they are relevant to IPV applications (e.g., 50 lx to 1000 lx). Therefore, reporting IPV performance at 50 lx, 200 lx, and 1000 lx, as per IEC 62607-7-2, is the minimum for consistent interlaboratory comparisons as part of ISC- or RC-based benchmarking.

### 6. Discussion and Conclusions

Our study comprehensively examined IPV characterization and benchmarking challenges, providing novel quantitative and theoretical insights as well as practical solutions. Our extensive experimental analyses of the *impact* of various error sources on IPV performance parameters enabled us to uncover the considerable PCE characterization errors (20% - 60%) due to the angular interplay among indoor TLS, MD, and DUT under diffuse IPV illumination. This issue—overlooked in IEC 62607-7-2 and the prior IPV literature—leads to a marked dependence of the apparent PCE on the working distance between an IPV and a diffuse indoor TLS. It would conceptually be necessary to thoroughly quantify—through experiment—the angular behavior of the IPV, MD, and indoor TLS across all angles to address this effect, which would be particularly time-consuming and would require specialized equipment not commonly available in IPV research laboratories. Consequently, this effect would hinder the standardization of IPV measurements, as the wide diversity in angular behavior of commercial indoor TLSs introduces too much variability to establish consistent illumination conditions across different IPV laboratories. To mitigate the IPV characterization errors resulting from this angular interplay, we proposed and validated approaches such as the black-tube and RC methods. In general, however, our investigation revealed that the most reliable way to overcome the impact of the TLS-MD-IPV angular interplay on IPV characterization is to use collimated or quasi-collimated indoor TLSs, which is therefore strongly recommended for the standardization of IPV characterization. Based on our experimental findings, a critical validation test for collimated IPV illumination involves measuring the IPV's photovoltaic performance at working distances with a ratio of at least 1:2, ensuring that apparent PCE changes are ≤5%.

From a benchmarking perspective, we quantitatively revealed, for the first time, that an indoor TLS's CCT and indoor spectral coincidence class do not provide unambiguous measures of benchmarking accuracy. We ascertained that these metrics may lead to PCE benchmarking discrepancies greater than the essential 5% accuracy threshold even for indoor TLSs with class-SA indoor spectral coincidence and greater than 20% for indoor TLSs with class-A indoor spectral coincidence. Crucially, we revealed that the interplay between the IPV's spectral response and the reference and test irradiance spectra may exacerbate the benchmarking error, while making it impossible to mitigate it through a single TLS-specific correction factor. Therefore, in the context of benchmarking approaches relying on the indoor spectral coincidence concept, we introduced and validated additional metrics to accurately assess the benchmarking error specific to the IPV, its operating condition, and its interplay with the reference and test irradiance spectra. If a TLS's indoor spectral coincidence is relied upon for IPV benchmarking, reporting these additional metrics is strongly recommended for accuracy.

On the other hand, we quantitatively demonstrated how the RC method can seamlessly and accurately enable the benchmarking of IPVs covering a broad performance parameter space against diverse reference IPV illumination conditions, while also exhibiting a high degree of insensitivity to diffuse IPV illumination.

These conclusions and guidelines, grounded in detailed error analyses, address crucial gaps in existing knowledge concerning IPV characterization, reporting, and benchmarking, providing practical, actionable solutions and leading to the extensive checklists in Supplementary Tables 6-12. These checklists are designed to achieve reliable IPV characterization and benchmarking with a target inaccuracy of ≤5%, aiming to facilitate meaningful comparisons of IPV performance across studies and thus advance the field effectively.

## Methods

*Indoor Test Light Sources*

The indoor TLSs included a BLPAN303065 LED panel (10 W, Biard, CCT6500K, 30 cm x 30 cm) driven by an adjustable driver and a collimated white LED spotlight (PLS-3000-030-12-S, Mightex, CCT3000K) with a computer-controlled driver providing 12-bit resolution for constant current output.

*Measuring Devices*

The ISS comprised a Firefly 4000 spectrometer (200 nm - 850 nm, Changchun New Industries Optoelectronics Technology) connected to an integrating sphere (2P3, Ø50 mm, Thorlabs) via a fiber adapter and patch cable (Thorlabs). It was radiometrically calibrated (±1.8% accuracy) under normal collimated illumination by NIST on March 2, 2023 (365 nm - 750 nm). The OPM comprised a calibrated photodiode (S120VC, 200 nm - 1100 nm, Thorlabs) and a PM400 power meter (Thorlabs) with NIST-traceable calibration (±5% accuracy) performed by Thorlabs on October 10, 2022. The RC (ID#NIST1014), supplied and calibrated by NIST (±0.37% accuracy) under normal collimated illumination, comprised a silicon solar cell with a KG-5 glass window. The luxmeters were an ALX-1309 (ATP Instrumentation, ±5% rdg ±10d accuracy; LM1) and an R8140 (REED, ±3% rdg ±5d accuracy; LM2), where 'rdg' and 'd' denote readings and digits, respectively.

*Devices Under Test*

M1 is a c-Si device (OSD15-5T, Centronic, 0.15 cm$^2$). M2 is an a-Si:H device (AM-1456CA-DGK-E, Amorton, Panasonic, 0.355 cm$^2$ per cell). H and L use perovskite and organic active layers, respectively (Supplementary Note 18). None were optimized for high-efficiency IPV. Spatial uniformity was assessed with a c-Si device (OSD1-5T, Centronic, 0.01 cm$^2$). Note that device areas do not affect the error analyses presented throughout our study, as these rely on relative PCE changes (cf., absolute PCE values) for a given device when the condition at hand is varied (Supplementary Note 19).

*Dark-Tube Method*

The dark-tube method relied on a black ABS Cell Core pipe (ASTM F628, IPEX, 40 mm inner diameter). Tubes of 4 cm and 12 cm lengths were cut to bridge the distance between the TLS and the DUTs/MDs, with DUTs/MDs placed at the tube center.

*Reference-Cell Method for IPV Characterization*

The RC method (Supplementary Note 11) relied on the ID#NIST1014 RC. TLS spectra were acquired with our calibrated ISS, ensuring accurate background subtraction. Spectral responsivity was measured using a PTS-2-QE Quantum Efficiency System (Sciencetech), calibrated with a broadband pyroelectric detector (SCI425CMA, Sciencetech, April 25, 2023). Spectral responsivity was measured in direct-current mode with a 10 nm wavelength interval. The illumination condition was fine-tuned by adjusting the TLS power, with continuous monitoring performed via a Python code controlling a Keithley 2614b source meter (Tektronix).

*Current-Voltage Measurements*

Current-voltage characteristics were measured using a Keithley 2614b source meter (Tektronix), sweeping the bias voltage from +1.0 V to -0.2 V at 40 mV/s. Measurements were performed in a dark, non-reflective enclosure in ambient air, with the devices under test mounted in dark, non-reflective fixtures. During J−V measurements under illumination, TLSs were securely mounted using optical posts and MDs/DUTs were accurately positioned within uniformly illuminated regions. TLS spatial uniformity was assessed by fixing the TLS and moving the MD (OSD1-5T) using a two-axis linear translation stage (Thorlabs).

*NIST Disclaimer*

Certain commercial equipment, instruments, software, or materials, commercial or non-commercial, are identified in this paper in order to specify the experimental procedure adequately. Such identification does not imply recommendation or endorsement of any product or service by NIST, nor does it imply that the materials or equipment identified are necessarily the best available for the purpose.

**Competing Interests Statement**

Irina Valitova works for a company commercializing perovskite photovoltaics. Sebastian B. Meier works for a company commercializing organic photovoltaics. Those companies did not fund this work. All other authors declare no competing interests.

**Data Availability Statement**

The authors declare that the data supporting the findings of this study are available within the paper and its Supplementary Information. Should any raw data files be needed in another format, they are available from the corresponding author upon reasonable request.